\documentclass[aps,prd,preprint,superscriptaddress,tightenlines,nofootinbib,showpacs]{revtex4}

\usepackage{graphicx}
\usepackage{dcolumn}
\usepackage{bm}

\def\mtiny{\vrule width 0pt}
\def\mrm#1{\mathrm{#1}}
\def\DZ{\relax\ifmmode{D^0}\else{$\mrm{D}^{\mrm{0}}$}\fi}
\def\DONE{\relax\ifmmode{D_1}\else{$\mrm{D}_{\mrm{1}}$}\fi}
\def\DTWO{\relax\ifmmode{D_2}\else{$\mrm{D}_{\mrm{2}}$}\fi}
\def\KZ{\relax\ifmmode{K^0}\else{$\mrm{K}^{\mrm{0}}$}\fi}
\def\KSHO{\relax\ifmmode{K_{\rm S}}\else{$\mrm{K}_{\mrm{S}}$}\fi}
\def\KLON{\relax\ifmmode{K_{\rm L}}\else{$\mrm{K}_{\mrm{L}}$}\fi}
\def\BZ{\relax\ifmmode{B^0_d}\else{$\mrm{B}^{\mrm{0}_d}$}\fi}
\def\BZp{\relax\ifmmode{B^0}\else{$\mrm{B}^{\mrm{0}}$}\fi}
\def\BZS{\relax\ifmmode{B^0_s}\else{$\mrm{B}^{\mrm{0}_s}$}\fi}
\def\DZS{\relax\ifmmode{D^{*+}}\else{$\mrm{D}^{\mrm{*+}}$}\fi}
\def\DZB{\relax\ifmmode{\overline{D}\mtiny^0}
        \else{$\overline{\mrm{D}}\mtiny^{\mrm{0}}$}\fi}\def\KZB{\relax\ifmmode{\overline{K}\mtiny^0}
        \else{$\overline{\mrm{K}}\mtiny^{\mrm{0}}$}\fi}
\def\BZB{\relax\ifmmode{\overline{B}\mtiny^0_d}
        \else{$\overline{\mrm{B}}\mtiny^{\mrm{0}_d}$}\fi}
\def\BZBp{\relax\ifmmode{\overline{B}\mtiny^0}
        \else{$\overline{\mrm{B}}\mtiny^{\mrm{0}}$}\fi}
\def\BZBS{\relax\ifmmode{\overline{B}\mtiny^0_s}
        \else{$\overline{\mrm{B}}\mtiny^{\mrm{0}_s}$}\fi}
\def\DZC{\relax\ifmmode{\overline{D}\mtiny^0}
        \else{$\overline{\mrm{D}}\mtiny^{\mrm{0}}$}\fi}

\begin{document}

\preprint{CLNS 05/1908}
\preprint{CLEO 05-3}         

\title{Search for $\DZ\!-\!\DZB$ Mixing in the Dalitz Plot Analysis of $\DZ \to K_S^0\pi^+\pi^-$}  


\setcounter{footnote}{0}

\date{March 19, 2005}

\begin{abstract} 
The resonant substructure in $\DZ \to
K^0_S\pi^+\pi^-$ decays
is described by a combination of ten
quasi two-body intermediate states which include
both $CP$-even and $CP$-odd eigenstates and one doubly-Cabibbo suppressed channel.
We present a formalism that connects the
variation in $\DZ$ decay time over the Dalitz plot
with the
mixing parameters, $x$ and $y$, that 
describe off-shell and on-shell  
$\DZ\!\!-\!\overline{D}\vphantom{D}^0$ mixing. 
We analyze the CLEO~II.V data sample and find the parameters $x$ and
$y$
are consistent with zero.
We limit $(-4.7\!<\!x\!<\!8.6)\%$ and
$(-6.1\!<\!y\!<\!3.5)\%$
at the 95\% confidence level.
\end{abstract}

{
\renewcommand{\thefootnote}{\fnsymbol{footnote}}

\author{D.~M.~Asner}
\author{S.~A.~Dytman}
\author{W.~Love}
\author{S.~Mehrabyan}
\author{J.~A.~Mueller}
\author{V.~Savinov}
\affiliation{University of Pittsburgh, Pittsburgh, Pennsylvania 15260}
\author{Z.~Li}
\author{A.~Lopez}
\author{H.~Mendez}
\author{J.~Ramirez}
\affiliation{University of Puerto Rico, Mayaguez, Puerto Rico 00681}
% ...
\collaboration{CLEO Collaboration} 
\noaffiliation

\maketitle

Studies of the evolution of a $\KZ$ or $\BZp$ into the respective
anti-particle, a $\KZB$ or $\BZBp$,
have guided the form and content
of the Standard Model and permitted
useful estimates of the 
masses of the charm~\cite{goodetal+glr} 
and top quark~\cite{Albrecht+rosner87} prior to 
their direct observation.
A $\DZ$ can evolve into a $\DZB$ through on-shell intermediate
states, such as $K^+K^-$ with mass, $m_{K^+K^-}\!=\!m_{\DZ}$, or through
off-shell intermediate states, such as those that might be present
due to new physics.  
This evolution through the former (latter) states is parametrized by
the dimensionless variables $-iy$ $(x)$ defined in Eq.~\ref{eqn:xandy}.

Many predictions for $x$ in the $\DZ\!\to\!\DZB$ amplitude have
been made~\cite{hnncomp}.  
Several non-Standard Models
predict $|x| > 0.01$.  Contributions
to $x$ at this level could result from
the presence of new particles with masses as high as 100-1000~TeV~\cite{lns+ark}.
The Standard Model short-distance contribution to $x$ is determined by
the box diagram in which two virtual quarks and two virtual $W$ bosons
are exchanged. The magnitude of $x$ is determined by the mass and 
Cabibbo-Kobayashi-Maskawa (CKM)~\cite{ckm} couplings of the virtual
quarks.
From the Wolfenstein parameterization~\cite{wolf} where
$\lambda\equiv\sin^2\theta_C\approx 0.05$, contributions
involving $b$ quarks ($\sim$$\lambda^6$) can be neglected relative to
those with $d$ and $s$ quarks ($\sim$$\lambda^2$). The most prominent
remaining amplitude is proportional to $(m^2_s - m^2_d)/m^2_W$. The near
degeneracy on the $W$ mass scale of the $d$ and $s$ quarks results in
a particularly effective suppression by the GIM\cite{gimktwz} mechanism.
A simple estimate of $x$ is obtained by comparing with the Kaon
sector;
\begin{equation}
\frac{\Delta M_\DZ}{\Delta M_\KZ}=\frac{f_\DZ (m^2_s - m^2_d)
m_\DZ}{f_\KZ (m^2_c - m^2_u) m_\KZ}\,.
\end{equation}
Assuming $f_\DZ \approx f_\KZ$ and taking $m_u = 5$~MeV, $m_d =
9$~MeV, $m_s = 60-170$~MeV, $m_c = 1.2$~GeV and $\Delta M_\KZ =
(3.48\pm0.01)\times 10^{-15}$~GeV, and $x = \frac{\Delta
M_\DZ}{\Gamma} = 6.31\times 10^{11}\times \Delta M_\DZ$ 
yields, $x = 2\times 10^{-5}\hbox{--}2\times 10^{-4}$.
Short distance contributions to $y$ are expected to be less than $x$. Both are beyond
current experimental sensitivity. Long distance effects are expected
to be larger but are difficult to estimate due to the large number of
resonances near the $\DZ$ pole. It is likely that $x$ and $y$
contribute similarly to mixing in the Standard Model.
Decisive signatures of new physics
include
$|y|\ll|x|$ or Type~II or Type~III $CP$ violation \cite{pdgcpv}.
In order to assess the origin
of a $\DZ\!-\!\DZB$ mixing signal, 
the values of both $x$ and $y$ must be measured.

Previous attempts to measure $x$ and $y$ include: the measurement of the
wrong sign semileptonic branching ratio
$\DZ\!\to\!K\ell\nu$~\cite{babarsl}
which is sensitive to the mixing rate $R_M=\frac{x^2+y^2}{2}$; 
decay rates to $CP$ eigenstates
$\DZ\!\to\!K^+K^-,\pi^+\pi^-$~\cite{Belley} which
are sensitive to $y$;
and the wrong sign $\DZ\!\to\!
K^+\pi^-$~\cite{cleokpi,BABARkpi,bellekpi}
hadronic branching ratio which measures 
$x^{\prime 2}=(y\sin\delta_{K\pi}\!+\!
x\cos\delta_{K\pi})^2$ and $y^\prime=y\cos\delta_{K\pi} -
x\sin\delta_{K\pi}$.
Here, $\delta_{K\pi}$, which has yet to be measured experimentally,
is the relative strong phase between
$\DZ$ and $\DZB$ to $K^+\pi^-$.
In this study we utilize the fact that the values of
$x$ and $y$ can also be determined from the distribution of the
$\DZ\to K^0_S\pi^+\pi^-$ Dalitz plot if one measures that distribution
as a function of the $\DZ$
decay time.  We show that $x$ and $y$ can be separately detemined.  This is
the first demonstration of possible sensitivity to the sign of $x$.
Predictions of the sign of $x$ are sensitive to the details of 
the treatment of long distance effects within the Standard Model 
as well as the nature of potential new physics contributions.

The time evolution of the $D{^0}\hbox{--}\overline D{^0}$ system
is 
described by the Schr\"odinger equation
\begin{equation}
i \frac{\partial}{\partial t} 
 {D{^0}(t)\choose \overline D{^0}(t)} = 
\left({\hbox{\bf M}} - \frac{i}{2} {{\bf \Gamma}}\right) 
 {D{^0}(t)\choose \overline D{^0}(t)} \, ,
\label{eqn:schro}
\end{equation}
where the {\bf M} and ${\bf \Gamma}$ matrices are Hermitian, 
and $CPT$ invariance requires 
$M_{11}=M_{22}\equiv M$ and $\Gamma_{11}=\Gamma_{22}\equiv \Gamma$.
The off-diagonal elements of these matrices
describe the dispersive or long-distance and absorptive
or short-distance contributions to 
$D{^0}\hbox{--}\overline D{^0}$ mixing.

The two eigenstates $D_1$ and $D_2$ of the effective Hamiltonian matrix 
$(\hbox{\bf M} - \frac{i}{2} {\bf \Gamma})$ 
are given by
\begin{equation}
|D_{1,2}\rangle = p | D{^0}\rangle \pm q |\overline D{^0}\rangle\;,
\;\;\;\; p^2+q^2 = 1\,.
\label{eqn:eigenstates}
\end{equation} 
The corresponding eigenvalues are
\begin{equation}
\lambda_{1,2}\!\equiv\!m_{1,2}\!-\!\frac{i}{2} \Gamma_{1,2}\!=\!\left(M\!-\!\frac{i}{2} \Gamma \right)
\!\pm\!\frac{q}{p} \left(M_{12}\!-\!\frac{i}{2} \Gamma_{12} \right),
\label{eqn:eigenvalues}
\end{equation}
where $m_{1,2}$, $\Gamma_{1,2}$ are the masses and decay widths
and
\begin{equation}
\frac{q}{p} = 
\sqrt{\frac{M^*_{12}-\frac{i}{2}\Gamma^*_{12}}{M_{12}-\frac{i}{2}
\Gamma_{12}}} 
\,.
\label{alpha}
\end{equation}
The proper time evolution of the eigenstates of Eq.~\ref{eqn:schro} 
is 
\begin{equation}
|D_{1,2}(t)\rangle = e_{1,2}(t)|D_{1,2}\rangle,\,e_{1,2}(t)=e^{[-i(m_{1,2}-\frac{i\Gamma_{1,2}}{2})t]}.
\end{equation}
A state that
is prepared as a flavor eigenstate $|\DZ\rangle$ or
$|\DZB\rangle$ at
$t=0$ will evolve according to
\begin{eqnarray}
 |\DZ(t)\rangle \!
=\frac{1}{2p}\!\left[p(e_1\!(t){ +}e_2\!(t))|\DZ
\rangle{ +}q(e_1\!(t){ -}e_2\!(t))|\DZB\rangle\!\right]\hphantom{..} 
\label{eqn:initD0} \\
 |\DZB(t)\rangle \!
=\frac{1}{2q}\!\left[p(e_1\!(t){ -}e_2\!(t))|\DZ
\rangle{ +}q(e_1\!(t){ +}e_2\!(t))|\DZB\rangle\!\right]\,. 
\label{eqn:initD0B}
\end{eqnarray}

We parameterize the $K^0_s\pi^+\pi^-$ Dalitz plot following 
the methodology described in Ref.~\cite{dalitz,bergfeld} using the same sign
convention as Ref.~\cite{e687a+b,ourdalitz,ourcpvdalitz}. 
Now, however, we generalize to the case where the time-dependent state is a
mixture of $\DZ$ and $\DZB$ so the Dalitz Plot distribution depends also
on $x$ and $y$.
We express the amplitude for $\DZ$ to decay via the $j$-th quasi-two-body state as
$a_j e^{i\delta_j}{\cal A}^j_k$
where ${\cal A}^j_k = {\cal A}^j_k(m_{K^0_S\pi^-}^2,m_{\pi\pi}^2)$ is
the Breit-Wigner amplitude for resonance $j$ with spin $k$ described
in Ref.~\cite{bergfeld}. We denote the $CP$ conjugate amplitudes for
$\DZB$ as ${\cal \overline A}^j_k = {\cal \overline A}^j_k(m_{K^0_S\pi^+}^2,m_{\pi\pi}^2)$.

We begin our search for $\DZ\!-\!\DZB$ mixing in $\DZ\to K^0_S\pi^+\pi^-$ from the results of our standard
fit in Ref.~\cite{ourdalitz} which clearly observed the ten modes,
($K^{\ast-} \pi^+$,
${K}^{\ast}_{0}\!(1430)^- \pi^+$,
${K}^{\ast}_{2}\!(1430)^- \pi^+$,
${K}^{\ast}(1680)^- \pi^+$,
$K^0_S \rho$,
$K^0_S \omega$,
$K^0_S f_0(980)$,
$K^0_S f_2(1270)$,
$K^0_S f_0(1370)$,
and the ``wrong sign'' $K^{\ast+} \pi^-$) plus a small non-resonant component.

The decay rate to $K^0_S\pi^+\pi^-$ with ($m^2_{K^0_S\pi^-}$,
$m^2_{\pi^+\pi^-}$) at time $t$ of a particle tagged as $|\DZ\rangle$
at $t=0$ is
\begin{equation}
d\Gamma(m^2_{K\pi},m^2_{\pi\pi},t)\!=\!
\frac{1}{256\pi^3 M^3}
\!\left|{\cal M}\right|^2\!dm^2_{K\pi}dm^2_{\pi\pi},
\end{equation}
where the matrix element is defined as ${\cal M} = \langle f|{\cal
H}|i\rangle$.
We evaluate $|{\cal M}|^2$ where $|i\rangle$ is given by Eq.~\ref{eqn:initD0},
and
\hbox{$\langle f| = \langle K^0_S\pi^+\pi^-(m^2_{K^0_S\pi^-},m^2_{\pi^+\pi^-})|$}.

The decay channels can be collected into those which are $CP$-even or
$CP$-odd (with amplitudes $A_+$ or $A_-$) and to those which are $\DZ$ or $\DZB$
flavor eigenstates (with amplitudes $A_{f}$ or $\overline
A_{f}$);
\begin{eqnarray}
\langle f |{\cal H}|D_{+,-}\rangle & = & \sum a_j
e^{i\delta_j}{\cal A}^j_{k={\rm even,odd}}  =  A_{+,-} \\ 
\langle \overline f|{\cal H}|D_{+,-} \rangle & = &\sum a_j
 e^{i\delta_j}{\cal \overline A}^j_{k={\rm even,odd}}  = \overline A_{+,-} \\
\langle f|{\cal H}|D^0_f\rangle & = & \sum a_j
e^{i\delta_j}{\cal A}^j_k  =  
A_f \\
\langle \overline f|{\cal H}|\overline{D}\mtiny^0_{\overline f}\rangle & = & \sum \bar a_j  e^{i\bar
\delta_j}{\cal \overline
A}^j_k  =  \overline A_{\overline f}.
\end{eqnarray}
Dalitz plot analyses are sensitive only to relative phases
and amplitudes. As in Ref.~\cite{ourdalitz}, we fix $a_\rho=1, \delta_\rho=0$
and assume $a_j = \bar
a_j$, $ \delta_j = \bar \delta_j$.
In Ref.~\cite{ourcpvdalitz}, we considered $CP$ violation more generally
and allowed $a_j \neq \bar a_j$, $ \delta_j \neq \bar \delta_j$.

Collecting terms with similar time dependence we find
\begin{eqnarray}
\langle f|{\cal H}| \DZ(t)\rangle=& &\!\frac{1}{2p}(\langle f|{\cal
H}|D_1(t)\rangle +\langle f|{\cal H}|D_2(t)\rangle)\\ \nonumber
=& &\!\frac{1}{2p}(\langle f|{\cal H}|(p\DZ+q\DZB)\rangle e_1(t) \\
\nonumber &+&\langle f|{\cal
H}|(p\DZ-q\DZB)\rangle) e_2(t)\\  \nonumber
=& &\!\frac{1}{2p}([p(A_f\!+\!A_{\rm +}+\!A_{\rm -}){\rm +}q(\overline A_f\!+\!
\overline A_{\rm +}\!+\!
\overline A_{\rm -}\!)]e_1(t)\\ \nonumber &+& [p(A_f\!+\!A_{\rm +}+\!A_{\rm -}){\rm -}q(\overline
A_f\!+\!\overline A_{\rm +}+\!\overline A_{\rm -})]e_2(t))\\  \nonumber
=& &\!\frac{1}{2}[(1+\chi_f)A_f{\rm +}(1+\chi_+)A_{\rm
+}{\rm +}(1+\chi_-)A_{\rm
-}]e_1(t)\hphantom{.}\hfill \\  \nonumber
&{\rm +}&\!\frac{1}{2}[(1-\chi_f)A_f{\rm +}(1-\chi_+)A_{\rm
+}{\rm +}(1-\chi_-)A_{\rm
-}]e_2(t)\hphantom{.} \\  \nonumber
\equiv & & e_1(t) A_1 + e_2(t) A_2 \hfill \hphantom{a}\\ 
\langle \overline f|{\cal H}|\DZB(t)\rangle\!=& &\!\frac{1}{2q}(\langle f|{\cal H}|D_1(t)\rangle -\langle f|{\cal H}|D_2(t)\rangle) \\  \nonumber
=& & \!\frac{1}{2}[(1\!+\!\chi^{-1}_{\overline f})\overline A_{\overline f}{\rm
+}(1\!+\!\chi^{-1}_{+})\overline A_{\rm +}{\rm
-}(1\!+\!\chi^{-1}_{-})\overline A_{\rm -}]e_1(t)\hphantom{.} \\  \nonumber
&{\rm +}&\!\frac{1}{2}[(1\!-\!\chi^{\rm -1}_{\overline f})\overline
A_{\overline f}{\rm +}(1\!-\!\chi^{\rm -1}_{\rm -})\overline A_{\rm
-}{\rm -}(1\!-\!\chi^{\rm -1}_{\rm -})\overline A_{\rm
-}]e_2(t)\hphantom{.} \\  \nonumber
\equiv & & e_1(t) \overline A_1 + e_2(t) \overline A_2 \,,
\hfill\hphantom{a}  \nonumber
\label{eqn:ampl}
\end{eqnarray}
for $\DZ$ and $\DZB$, respectively. Similar to Ref.~\cite{ampli},
\begin{eqnarray}
\chi_f \;  = & \hphantom{\pm} \frac{q}{p} \frac{\overline
A_{f}}{A_{f}} &  = \hphantom{\pm}  \left|\frac{\overline A_{f}}{A_{f}}\right|
\frac{1-\epsilon}{1+\epsilon}e^{i(\delta+\phi)} \,\hphantom{,}\\
\chi_{\overline f} \;  = & \hphantom{\pm} \frac{q}{p} \frac{\overline
A_{\overline f}}{A_{\overline f}} &  = \hphantom{\pm}  \left|\frac{\overline A_{\overline
f}}{\overline A_{f}}\right|
\frac{1-\epsilon}{1+\epsilon}e^{-i(\delta-\phi)} \,\hphantom{,}\\
\chi_{\pm} \;  = & \frac{q}{p}\frac{\overline A_{\pm}}{A_{\pm}} & = \pm \frac{1-\epsilon}{1+\epsilon}
e^{i\phi}\,,
\end{eqnarray} 
where $\delta$ is the relative strong phase between $\DZ$ and $\DZB$
to $K^0_S\pi^+\pi^-$, and in the limit of
$CP$ conservation, the real $CP$-violating parameters, $\epsilon$ and $\phi$, are zero.
Squaring the amplitude and factoring out the time dependence yields
\begin{eqnarray}
\left|{\cal M}\right|^2  & = &\left|e_1 (t) \right|^2 \left|A_1 \right|^2 
  { +}  \left| e_2 (t) \right|^2\left| A_2 \right|^2 \\ \nonumber
 & &+ 2\Re[e_1 (t) e_2^* (t)A_1 A^*_2]
\,\hphantom{.} \label{eqn:msq} \\
\left|{\cal \overline M}\right|^2 & = & \left| e_1 (t) \right|^2 \left| \overline A_1 \right|^2 
  +  \left| e_2 (t) \right|^2\left| \overline A_2 \right|^2 \\ \nonumber
 & &+ 2\Re[e_1 (t) e^*_2 (t)\overline A_1 \overline A^*_2]. \label{eqn:mbarsq}
\end{eqnarray}
The time-dependent terms are given explicitly by
\begin{eqnarray}
 \left|e_{1,2}(t)\right|^2 & = &
\exp {\left(2\Im(\lambda_{1,2})t\right)}{ =}\exp {\left( { -} \Gamma_{1,2}
t\right)}\\ \nonumber & = &\exp {\left( { -} \Gamma
(1\pm y)t\right)} \\
e_1(t)e_2(t)^* & = & \exp {\left( { -} i\lambda_1t\right)}\exp {\left(
{ +} i\lambda_2t\right)} \\ \nonumber
& = &\exp {\left( { -} \Gamma(1 { +} ix)t\right)} 
\end{eqnarray}
where
\begin{equation}
\Gamma =\frac{\Gamma_1\!+\!\Gamma_2}{2}\,, \;\;\;\;\;
x =  \frac{m_1\!-\!m_2}{\Gamma}\,, \;\;\;\;\;
y = \frac{\Gamma_1\!-\!\Gamma_2}{2\Gamma}\,.
\label{eqn:xandy}
\end{equation}
Experimentally, $y$ modifies the lifetime of certain contributions
to the Dalitz plot while $x$ introduces a sinusoidal rate variation

This analysis uses an integrated luminosity of 9.0~fb$^{-1}$
of $e^+e^-$ collisions at $\sqrt{s}\approx10\,$GeV provided by
the Cornell Electron Storage Ring (CESR).
The data were taken with the CLEO~II.V 
detector~\cite{ctwo}. 
The event selection is identical to that used in our previous study
of $\DZ \to K^0_S \pi^+\pi^-$~\cite{ourdalitz,ourcpvdalitz} which did not
consider $\DZ\!-\!\DZB$ mixing.
We reconstruct candidates for the decay sequence
$D^{\ast+}\!\to\!\pi^+_S \DZ$, $\DZ\!\to\!K^0_S\pi^+\pi^-$.
The charge of the slow
pion ($\pi^+_S$ or $\pi^-_S$) identifies the initial charm state 
as either $\DZ$ or $\DZB$. The detector resolution in the 
Dalitz plot parameters $m^2_{K\pi}$ and $m^2_{\pi\pi}$ is small relative
to the intrinsic widths of intermediate resonances; the exception
is the decay channel $\DZ\!\to\!K^0_S\omega, \omega\!\to\!\pi^+\pi^-$. 
We reconstruct the $\DZ$ decay time $t$ as described in Ref.~\cite{cleokpi}.

The uncertainty in $t$, $\sigma_t$, is typically 200~fs or
$0.5/\Gamma$ and cannot be neglected.
We fit the unbinned decay time distribution by analytically convolving the exponentials
in each term in Eqs.~\ref{eqn:msq} and~\ref{eqn:mbarsq}
by a
resolution function similar to, but
slightly modified from, that used in Ref.~\cite{CLEOy} and
Ref.~\cite{dlife}.
The signal likelihood is represented as the sum of an exponential
convolved with two Gaussians.  The width of the first Gaussian is the
event-by-event measured proper time error, $\sigma_{t}$, times a
scale factor, $S_{\rm Sig}$, which allows
for a uniform mistake in the covariance matrix elements
of the $D$ meson and its daughters, perhaps due to an imperfect
material description of the detector during track fitting.  
For the other Gaussian, the measured proper time
errors are ignored and the width $\sigma_{\rm MisSig}$ and
the normalization $f_{\rm MisSig}$ are fit for directly.  
This Gaussian models the `MIS-measured SIGnal'' proper time resolution when 
the measured $\sigma_{t}$ is {\em not} correct,
as would be the case for hard multiple scattering of one
or more of the $D$ meson daughters.  The sum of these
two components to the likelihood is normalized by the total signal
fraction $f_{\rm sig}$. 
Note that if we understand our
detector well, we will find that the scale factor used in the first
Gaussian is close to unity and the fraction of the signal in the second
Gaussian is near zero.
\begin{figure}[t]
\includegraphics*[width=.99\textwidth]{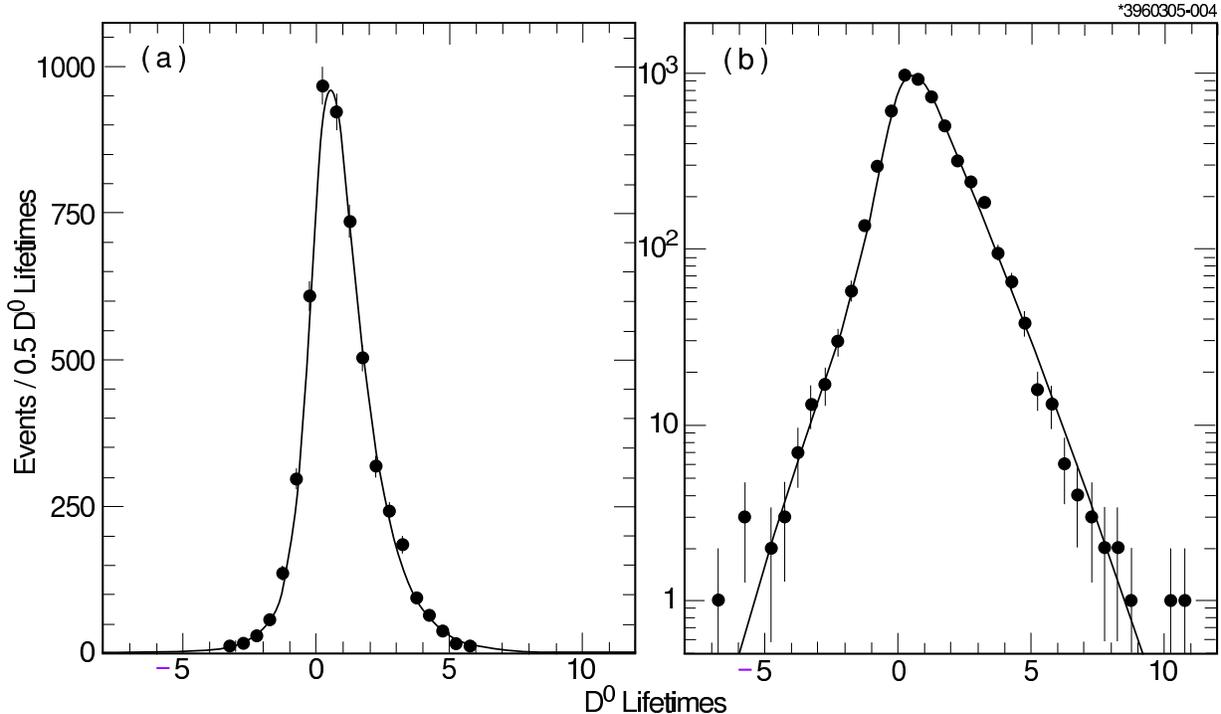}
\caption
{\label{fig:t}
Projection of the results of Fit~A onto the $\DZ$ decay time with a) linear and b)
logarithmic vertical scale.
}
\end{figure}

The treatment of the background is similar to that of the signal.  The
total background likelihood is normalized by the background
fraction, which is ($1 - f_{\rm sig}$).  We consider
two types of background: background with zero lifetime
and background with non-zero lifetime $\tau_{\rm BG}$ normalized by 
$f_{\tau_{\rm BG}}$.  We constrain both
backgrounds to have the same resolution function.  The model
for the resolution function is two Gaussians, with core width
$\sigma_{\rm BG}$, misreconstructed width $\sigma_{\rm MisBG}$ and
the background fraction $f_{\rm MisBG}$ in the wider Gaussian.

We perform an unbinned maximum likelihood fit to the Dalitz plot
which minimizes the
function ${\cal F}$ given below
\begin{equation}
{\cal F} = \sum_{\DZ}-2\ln {\cal
L}+\sum_{{\DZB}}-2\ln {\cal \overline
L}\,,
\label{eqn:like}
\end{equation}
where ${\cal L}$ and ${\cal \overline L}$ are defined as in
Ref.~\cite{ourcpvdalitz} using ${\cal M}$ and
${\cal \overline M}$ as defined in Eqs.~\ref{eqn:msq}
and~\ref{eqn:mbarsq} convolved with the resolution function described above.
Simplified Monte Carlo studies indicate that our fit procedure is unbiased and
the statistical errors as determined by the fit are accurate.

Our standard fit to the data, described above, 
is referred to as Fit~A. 
Fit~B is identical to Fit~A except $CP$ conservation ($\epsilon=0$, $\phi=0$)
is assumed.
The $\DZ$ and $\DZB$ sub-samples are fit independently in Fit~C1 and
Fit~C2, respectively. Fit~C1 and Fit~C2 are identical to Fit~B.

Fit~A has 35 free parameters;
ten resonances and the non-resonant contribution correspond to ten relative
amplitudes and ten relative phases, signal fraction and
mis-tag fraction, four signal decay time parameters, five background
decay time parameters, two mixing parameters and two $CP$-violating
parameters.
The results for $x$, $y$, $\epsilon$ and $\phi$ are in
Table~\ref{tab:results}
and are consistent with the absence of both $\DZ\!-\!\DZB$ mixing and
$CP$ violation.
The one-dimensional, 95\% confidence intervals
are determined by
an increase in negative log likelihood ($-2\ln{\cal L}$) of 3.84~units.
All other fit variables
are allowed to vary to distinct, best-fit
values.
The amplitude and phase, $a_j$ and $\delta_j$, for all fits 
in Table~\ref{tab:results},
are consistent with 
our ``no mixing'' result~\cite{ourdalitz}.
The projection of the results of Fit~A onto the $\DZ$ decay time is shown in Fig~\ref{fig:t}.
\begin{table}
\caption{
Results of the Dalitz-plot vs decay time fit of the
$\DZ\!\to\!K^0_S\pi^+\pi^-$.
Fit A allows
both $\DZ\!-\!\DZB$ mixing and $CP$ violation. 
Fit B is the $CP$-conserving fit, $\epsilon=0$ and $\phi=0$.
Fit C1 (C2) is the fit to the $\DZ$ ($\DZB$) sub-sample.
The errors shown for Fit~A and Fit~B are statistical, experimental systematic
and modeling systematic respectively and the 95\% confidence intervals
\emph{include} systematic uncertainty. The errors for Fit~C1
and Fit~C2 are statistical only.}
\label{tab:results}
\begin{tabular}{clc}
{Parameter} & {Best Fit} & {1-Dimensional 95\% C.L.}\\ \hline \hline
Fit A \hfill \hphantom{a} & \multicolumn{2}{c}{Most General Fit} \\
$x$ (\%) & $\hphantom{-}2.6^{+3.8}_{-9.0}\pm 0.4 \pm 0.4 $ &  $|x|<9.8\%$ \\
$y$ (\%) & $-0.3\!^{+4.0}_{-4.6} \pm 0.8 \pm 0.4 $ & $|y|<9.5\%$ \\ 
$\epsilon \hphantom{(\%)}$  & $-0.3\pm 0.5$ &              \\
$\phi$ ($^o$) & $\hphantom{-.}42\pm 78$  &  \\ \hline\hline
Fit B \hfill \hphantom{a} & \multicolumn{2}{c}{$CP$-conserving fit}  \\
$x$ (\%) & $\hphantom{-}1.9^{+3.2}_{-3.3}\pm 0.4 \pm 0.4 $ &  (-4.7:8.6) \\
$y$ (\%) & $-1.4\pm 2.4 \pm 0.8 \pm 0.4$ & (-6.1:3.5) \\ \hline\hline
Fit C1 \hfill \hphantom{a} & \multicolumn{2}{c}{$\DZ$ sub-sample}  \\
$x$ (\%) & $\hphantom{-}3.3^{+5.0}_{-4.8}$ &  (-6.1:13.5) \\
$y$ (\%) & $-2.8\!^{+3.6}_{-3.7}$ & (-10.2:4.2) \\ \hline
Fit C2 \hfill \hphantom{a} & \multicolumn{2}{c}{$\DZB$ sub-sample}  \\
$x$ (\%) & $\hphantom{-}0.6^{+5.7}_{-8.6}$ &  (-16.0:11.5) \\
$y$ (\%) & $-0.3\!^{+6.9}_{-3.1}$ & (-6.6:13.0) \\ \hline \hline
\end{tabular}
\end{table}

We find the parameters describing the signal decay time, $f_{\rm sig}=(97.1\pm0.8)\%$,
$\tau_{\rm sig}=402\pm 6$~fs, $S_{\rm Sig}=1.13\pm 0.02$, 
$\sigma_{\rm MisSig}=735\pm 155$~fs,
$(1-f_{\rm MisSig})=(96.9\pm 1.5)\%$ and the parameters describing the
background time, $f_{\tau_{\rm BG}}=(100\pm 1)\%$,
$\tau_{\rm bg}=94\pm 59$~fs, $(1-f_{\rm MisBG})=(87\pm 11)\%$, $\sigma_{\rm BG}=197\pm 39$~fs,
$\sigma_{\rm MisBG}=1116\pm 321$~fs. The scale factor $S_{\rm Sig}$,
although not consistent with unity, is comparable to results from
other CLEO lifetime analyses which include Ref.~\cite{CLEOy, cleokpi, dlife}.

We evaluate a contour in the two-dimensional
plane of $y$ versus $x$ that
contains the true value of $x$ and
$y$ at 95\% confidence level (C.L.)
without assumption regarding the relative strong
phase between $\DZ$ and $\DZB \to K^0_S\pi^+\pi^-$.
We determine the contour around our best-fit values where the $-2\ln{\cal L}$ has increased
by 5.99~units.  All fit variables other than $x$ and $y$  
are allowed to vary to distinct, best-fit
values at each point on the contour.
The contour for Fit~A is shown in Fig.~\ref{fig:xpyp}.
On the axes of $x$ and $y$, these contours
fall slightly outside the one-dimensional intervals listed in Table~\ref{tab:results},
as expected. 
The maximum excursion of the contour of Fit~A (Fit~B) from the origin
corresponds to a 95\% C.L. limit on the mixing rate of $R_M < 0.84\%$
($R_M < 0.55\%$).

We take the 
sample variance of $x$, $y$, $\epsilon$ and $\phi$
from the nominal result compared to the results in
the series of fits described below as a measure of the 
experimental systematic and modeling 
systematic uncertainty.
\begin{figure}[t]
\includegraphics*[width=.9\textwidth]{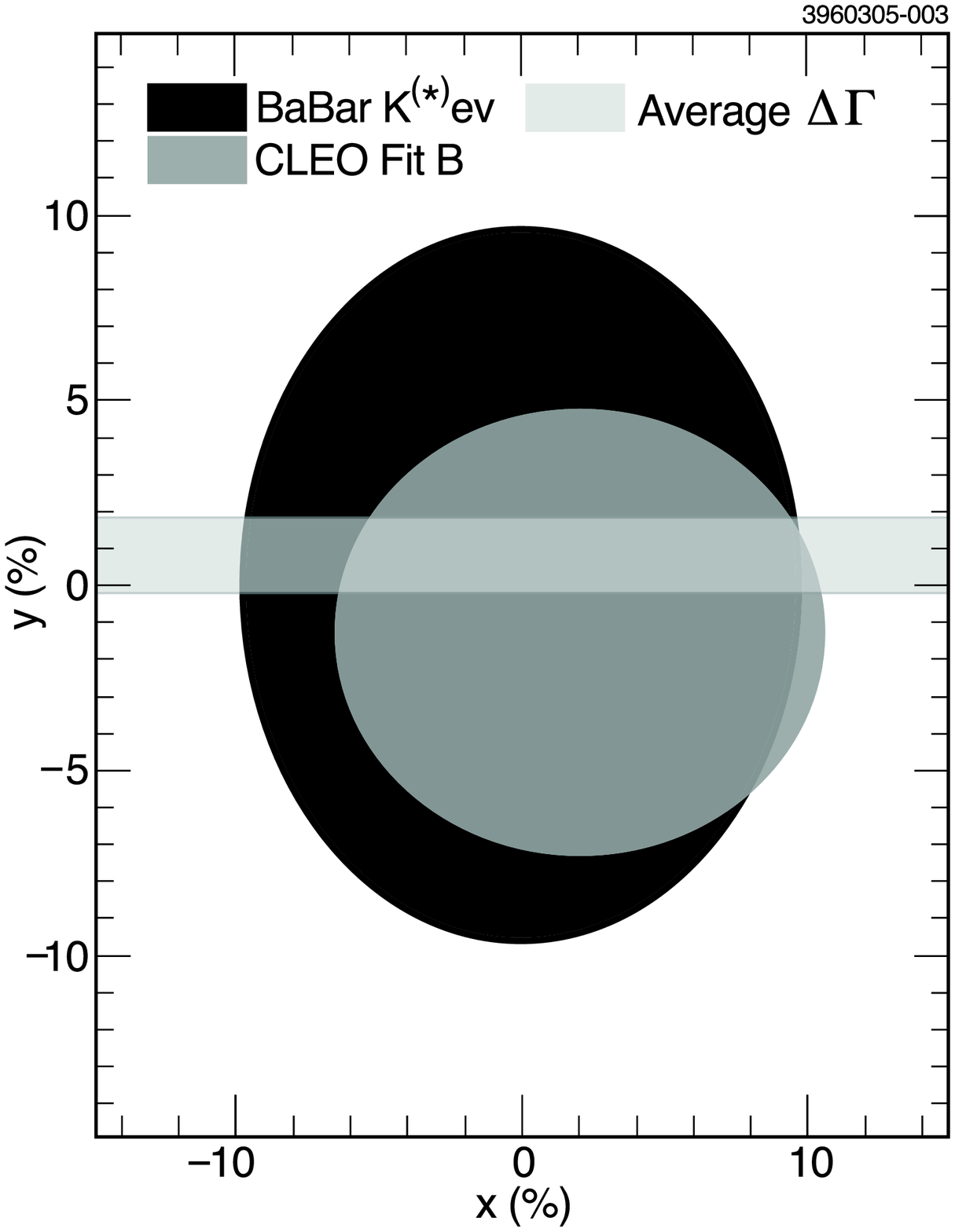}
\caption
{\label{fig:xpyp}
Allowed regions in the plane of $y$ versus $x$. No assumption
is made regarding $\delta_{K^0_S\pi^+\pi^-}$. The two-dimensional 95\%
allowed regions from our 
Fit~B (light shaded region) is shown. 
The allowed region for
$\Delta\Gamma$ is the average of the
$y_{CP}$~\cite{CLEOy,Belley}
results. 
Also shown is the limit from $\DZ\!\to\!K^{(*)}\ell\nu$ from
BABAR~\cite{babarsl}.
All results are consistent with the absence of mixing.
The limits from
CLEO~\cite{cleokpi} and BABAR~\cite{BABARkpi} from $D\to K\pi$
have similar sensitivity to Fit~B. The 95\% allowed regions (not shown) are circles of radius 5.8\% and 5.7\%, respectively,
when assumptions regarding $\delta_{K\pi}$ are removed.
The 95\% allowed region from Belle~\cite{bellekpi} also from $D\to
K\pi$ is more restrictive - a circle of radius 3.0\%.
}
\end{figure}

We consider systematic uncertainties from experimental sources and
from the decay model separately.
Our general procedure is to change some aspect of our fit and
interpret the change in the values of the mixing and $CP$-violating parameters in the non-standard fit relative to our nominal fit as an
estimate of the systematic uncertainty. 
Contributions to the experimental systematic uncertainties arise
from our 
model of the background, the efficiency, the event selection criteria,
and biases due to experimental
resolution as described in Ref.~\cite{ourdalitz}.
Additionally, we vary aspects of the decay time parametrization.
To estimate the systematic uncertainty regarding the $u \bar u, d \bar
d, s \bar s$ content of the background, we perform
fits where the background is forced to be all zero lifetime and all non-zero lifetime.
We consider a single or a triple rather than a double Gaussian to 
model the decay time resolution of the signal and background. 
We also vary by $\pm 1\sigma$ the fraction of misreconstructed signal $f_{\rm MisSig}$.
Finally, we set the scale factor for the measured proper time errors $S_{sig}$ to unity.
Variation in the event selection criteria are the largest contribution
to the experiment systematic error.

	Contributions to the theoretical systematic uncertainties
arise from our choices
for the decay model for $\DZ \to K_S^0\pi^+\pi^-$ as described in Ref.~\cite{ourdalitz}.  
We also consider the uncertainty
arising from our choice of resonances included in the fit.
To study the
stability of our results with other choices of resonances, we performed
fits which included additional resonances to the ones in our standard fit.
We compared the result of our nominal fit to a series of fits where
each of
the resonances, $\sigma$ or $f_0(600)$ and  $f_0(1500)$ which are
$CP$-even, and $\rho(1450)$ and $\rho(1700)$ which are $CP$-odd
 were included one at a time. 
In the standard fit we enumerate the non-resonant component with the
$K^*$ resonanaces.
We also considered fits where the non-resonant component was
considered to be $CP$-even or $CP$-odd.
Finally, we consider a fit that includes doubly-Cabibbo suppressed 
contributions from
$K_0(1430)$, $K_2(1430)$ and $K^*(1680)$ constrained to have the same
amplitude and phase relative to the corresponding Cabibbo favored
amplitude 
as the $K^*(892)$. There is no single dominant contribution to the modeling
systematic error.

In conclusion, we have analyzed the time dependence of the three-body
decay $\DZ \to K^0_S\pi^+\pi^-$ and exploited the interference between
intermediate states to limit the mixing parameters $x$ and
$y$ without sign or phase ambiguity.
Our data 
are consistent with an absence of both $\DZ\!-\!\DZB$ mixing and $CP$ violation. 
The two-dimensional limit in the 
mixing parameters, $x$ versus $y$, is similar to
previous results obtained from the same data
sample~\cite{cleokpi}, when assumptions regarding $\delta_{K\pi}$ are removed.
We limit $(-4.7\!<\!x\!<\!8.6)\%$ and
$(-6.1\!<\!y\!<\!3.5)\%$, at the 95\% C.L., with the assumption of
$CP$-conservation. We measure the $CP$-violating parameters
$\epsilon = -0.3 \pm 0.5$ 
and $\phi = 42^o \pm 78^o$.

We thank Alex Kagan, Yuval Grossman, and Yossi Nir for valuable discussions.
We gratefully acknowledge the effort of the CESR staff in providing us with
excellent luminosity and running conditions.
This work was supported by 
the National Science Foundation,
the U.S. Department of Energy, and the Natural Sciences and Engineering Council
of Canada.

\end{document}